\documentclass[aps,prl,twocolumn,superscriptaddress]{revtex4-1}

\usepackage{graphics}
\usepackage{graphicx}
\usepackage{color}
\usepackage{epstopdf}
\usepackage{amsmath}

\usepackage{cancel} 

\begin{document}

\title{Surface energy of strained amorphous solids } 

\author{Rafael D. Schulman}
\affiliation{Department of Physics and Astronomy, McMaster University, 1280 Main St. W, Hamilton, ON, L8S 4M1, Canada.}
\author{Miguel Trejo}
\affiliation{Laboratoire de Physico-Chimie Th\'eorique, UMR CNRS Gulliver 7083, ESPCI Paris, PSL Research University, 75005 Paris, France.}
\author{Thomas Salez}
\affiliation{Laboratoire de Physico-Chimie Th\'eorique, UMR CNRS Gulliver 7083, ESPCI Paris, PSL Research University, 75005 Paris, France.}
\affiliation{Univ. Bordeaux, CNRS, LOMA, UMR 5798, F-33405 Talence, France.}
\affiliation{Global Station for Soft Matter, Global Institution for Collaborative Research and Education, Hokkaido University, Sapporo, Hokkaido 060-0808, Japan.}
\author{Elie Rapha\"{e}l}
\affiliation{Laboratoire de Physico-Chimie Th\'eorique, UMR CNRS Gulliver 7083, ESPCI Paris, PSL Research University, 75005 Paris, France.}
\author{Kari Dalnoki-Veress}
\email{dalnoki@mcmaster.ca}
\affiliation{Department of Physics and Astronomy, McMaster University, 1280 Main St. W, Hamilton, ON, L8S 4M1, Canada.}
\affiliation{Laboratoire de Physico-Chimie Th\'eorique, UMR CNRS Gulliver 7083, ESPCI Paris, PSL Research University, 75005 Paris, France.}

\date{\today}


\begin{abstract}

Surface stress and surface energy are fundamental quantities which characterize the interface between two materials. Although these quantities are identical for interfaces involving only fluids, the \emph{Shuttleworth effect} demonstrates that this is not the case for most interfaces involving solids, since their surface energies change with strain. Crystalline materials are known to have strain dependent surface energies, but in amorphous materials, such as polymeric glasses and elastomers, the strain dependence is debated due to a dearth of direct measurements. Here, we utilize contact angle measurements on strained glassy and elastomeric solids to address this matter. We show conclusively that interfaces involving polymeric glasses exhibit strain dependent surface energies, and give strong evidence for the absence of such a dependence for incompressible elastomers. The results provide fundamental insight into our understanding of the interfaces of amorphous solids and their interaction with contacting liquids.
\end{abstract}

\maketitle

The surface energy $\gamma$ of an interface between two materials is the energetic cost associated with creating a unit of surface by cleaving and is associated with breaking intermolecular bonds, whereas the surface stress $\Upsilon$ characterizes the force required to generate a unit of area by deforming the materials and is associated with stretching the bonds of the molecules near the interface~\cite{Shuttleworth1950,Cammarata1994,Ibach1997,Sander2003,Muller2004}. These separate quantities are related through the Shuttleworth equation~\cite{Shuttleworth1950}, for which a simplified form is:

\begin{equation}
\Upsilon_{\textrm{AB}} (\epsilon)  = \gamma_{\textrm{AB}} (\epsilon) + \frac{\mathrm{d} \gamma_{\textrm{AB}}(\epsilon)}{\mathrm{d} \epsilon},
\label{Shuttleworth}
\end{equation}
where $\epsilon$ is the strain parallel to the interface and the subscript refers to the interface between A and B. Although the validity of Eq.~\ref{Shuttleworth} is well established (see~\cite{Cammarata1994} for a review), it is not obvious how the surface energy is dependent upon strain. For a liquid in contact with a vapour, the surface stress and energy are identical because as the liquid-vapour interface is deformed, molecules from the two fluids may simply rearrange themselves to maintain a constant average molecular environment at the interface. In contrast, as a crystalline solid is deformed, the surface density of atoms is altered, leading to a strain dependent surface energy. Given Eq.~\ref{Shuttleworth}, this strain dependence implies that the surface energy and surface stress are in general not equal for this class of materials. There has been some experimental work verifying this principle, but there is difficulty in performing absolute measurements which are precise and model independent~\cite{Nicolson1955,Cammarata1994,Ibach1997,Muller2004,Berger1997,Mays1968,Wasserman1970,Wasserman1972}. Due to the periodic structure of crystals, rigorous theoretical calculations of surface energy and stress are tractable and have been carried out for a multitude of these materials~\cite{Cammarata1994,Ibach1997,Muller2004,Nicolson1955}. 

On the other hand, there is little evidence to indicate whether the surface energies of interfaces involving amorphous materials, such as glasses and elastomers, are strain dependent. In fact, to the best of our knowledge, there has been no theoretical or experimental work investigating the Shuttleworth effect for glasses. Elastomers have recently received particular attention, because in these soft materials the surface stresses may induce large-scale deformations~\cite{Shanahan1987,Bostwick2014}. For instance, sufficiently soft cylindrical structures will undergo a Plateau-Rayleigh instability~\cite{Mora2010}. However, despite the multitude of recent studies, no consensus has been reached on whether interfaces involving elastomers have surface stresses which are different from surface energies ~\cite{Marchand2012,Nadermann2013,Style2013a,Weijs2014,Park2014,Mondal2015,Andreotti2016a,Andreotti2016,Xu2016,Xu2017,Xu2017b,Liang2018}. This situation is likely rooted in the fact that interpretations of the experimental measurements to determine $\Upsilon$ have been model dependent.

\begin{figure}[t!]
     \includegraphics[width=\columnwidth]{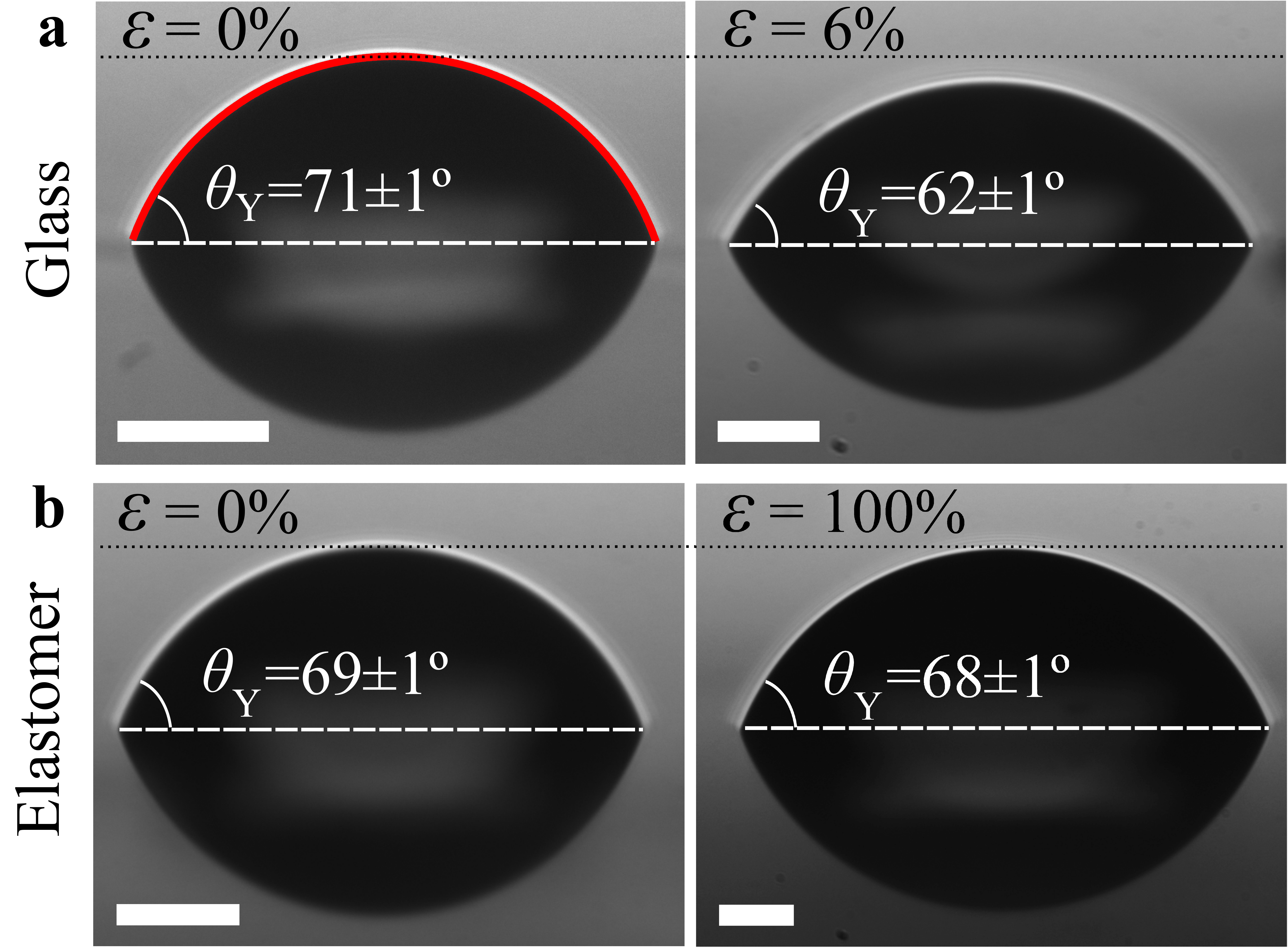}
\caption{(a) Contact angle measurements of glycerol on unstrained (left) and 6\,\%-strained (right) polycarbonate glass. The left panel shows a typical circular cap fit (red solid line) to the droplet profile. (b) Contact angle measurements of glycerol on unstrained (left) and 100\,\%-strained (right) Elastollan elastomer. The images are rescaled so that all contact radii appear equal, while preserving the aspect ratio. Since the profiles are spherical caps, the height of the cap is thus indicative of the contact angle (see the black dotted lines). Anything below the horizontal white dashed line is a reflection off the substrate. All scale bars correspond to 50 $\mu$m. }
\label{fig2}
\end{figure} 

In this study, by using contact angle measurements, we unambiguously quantify the strain dependence of the difference between the solid-liquid and solid-vapour surface energies of strained interfaces involving polymeric glassy and elastomeric materials. We employ Young-Dupr\'e's law, which dictates that $\gamma_\mathrm{lv} \mathrm{cos}\theta_\mathrm{Y} = \gamma_\mathrm{sv}-\gamma_\mathrm{sl}$, where s, l, and v indicate the solid, liquid, and vapour phases and $\theta_\mathrm{Y}$ is the contact angle at equilibrium. Since $\gamma_\mathrm{lv}$ is independent of any strain applied to the solid, $\theta_\mathrm{Y}$ is a direct indicator of the strain dependent difference between $\gamma_\mathrm{sv}$ and $\gamma_\mathrm{sl}$. We find that interfaces involving polymeric glassy materials do exhibit strain dependent surface energies. As seen in Fig.~\ref{fig2}(a), a droplet placed on a glassy substrate strained by only 6\,\% exhibits a significant change in contact angle. In contrast, we provide strong evidence that interfaces involving an elastomer together with a liquid or a vapour have surface energies which are unchanged by strain. In Fig.~\ref{fig2}(b), an elastomeric substrate strained by 100\,\% shows no measurable change in $\theta_\mathrm{Y}$. As we will show, $\theta_\mathrm{Y}$ is independent of strain for all tested combinations of elastomer and liquid. 

\section{Results}

\begin{figure}[t!]
     \includegraphics[width=\columnwidth]{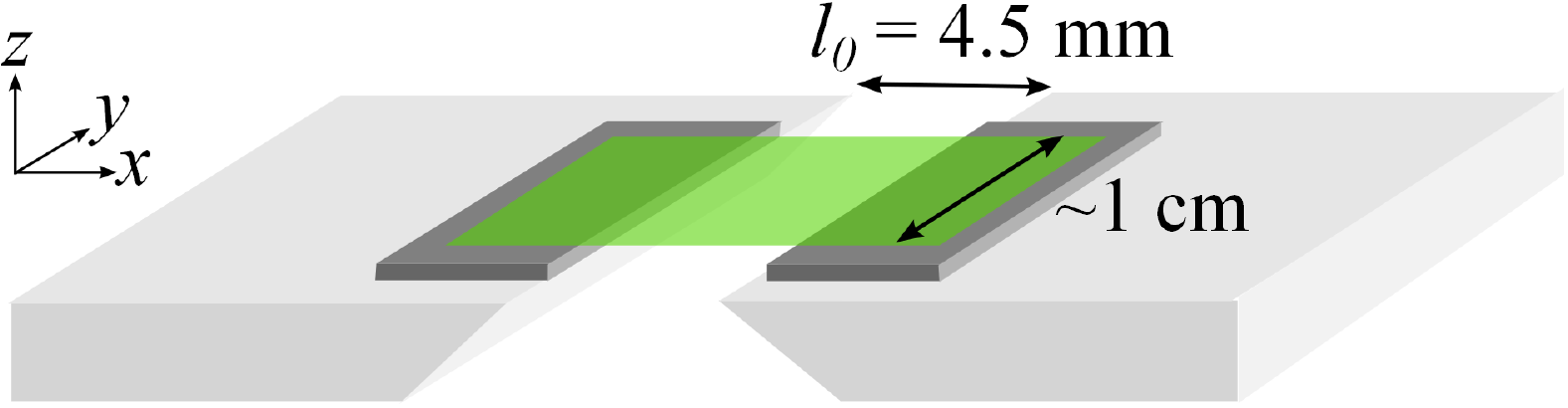}
\caption{Schematic of the sample holder used to apply precise strains to the films. The sample holder consists of two aluminum blocks separated by a fixed initial distance $l_0$. Two pieces of silicon which have been coated with the same polymeric
material as is to be strained, are affixed to the aluminum blocks. The films to be strained are placed such that they bridge the gap between the sample holder blocks. The strong adhesion between the film and the coated silicon pieces keeps the film in place and prevents delamination. The two blocks are then precisely separated by an additional distance $\Delta l$ along the $x$-direction, using a motorized translation stage at constant speed, which creates a strain $\epsilon = \Delta l/l_0$ in the film.}
\label{fig1}
\end{figure}

In our experiment, polymeric glassy and elastomeric films are strained (Fig.~\ref{fig1}) and then transferred onto a silicon wafer. We then place sub-millimetric liquid droplets on those strained films. The droplets are observed to be completely circular when viewed from above. We perform contact angle measurements by viewing the droplets from the side under an optical microscope and fitting their profiles to circular caps, an example of which  is shown in the left panel of Fig.~\ref{fig2}(a). We note that Young-Dupr\'e's law only holds for droplets which are much larger than the elastocapillary length of the system~\cite{Style2013a}. Therefore, we work exclusively in the regime where the droplet size much exceeds this length scale. All contact angle measurements are performed in air and at room temperature ($\sim 20^\circ$C). 

\subsection{Measurements on Polymeric Glasses}
In the first part of this study, we perform our measurements on polymeric glasses. Since we want to avoid any plastic deformation of the samples, we choose polysulfone (PSf) and polycarbonate (PC) which have large elongations at yield: up to 6\,\% and 8\,\% respectively for bulk samples~\cite{Margolis1985}. For the thin film samples and low strain rates employed, we find that both glasses can be strained up to 7-8\,\% without observing crazing. Thus, we vary the strains in the range of 0-8\,\% and discard any sample where crazing is observed. The absolute error in the strain is estimated to be $\pm$1\,\%. Contact angle measurements are performed with two standard test liquids: diiodomethane (DIM), a symmetric, non-polar molecule with $\gamma_\mathrm{lv} =  51$ mJ m$^{-2}$ (at $20^\circ$C)~\cite{Good1970}, and glycerol, a highly polar molecule with $\gamma_\mathrm{lv} $ = 63 mJ m$^{-2}$(at $20^\circ$C)~\cite{Lide2004}.

\begin{figure}[h!]
     \includegraphics[width=\columnwidth]{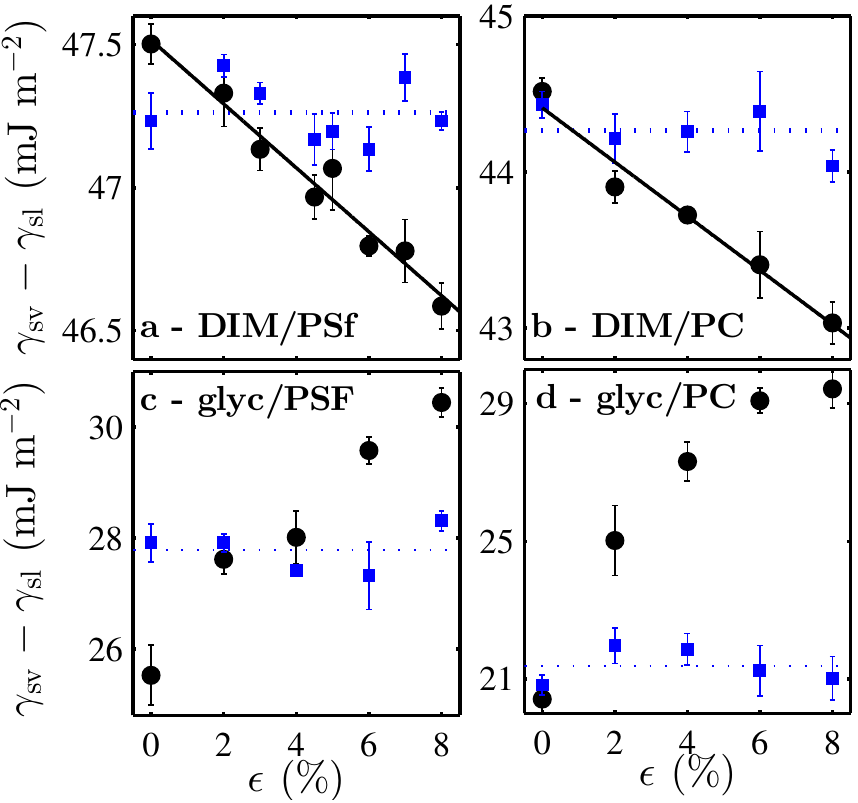}
\caption{Difference between solid-vapour and solid-liquid surface energies as a function of strain in the solid, shown as circle markers for four liquid/solid combination: (a) diiodomethane/polysulfone, (b) diiodomethane/polycarbonate, (c) glycerol/polysulfone, and (d) glycerol/polycarbonate. The square markers represent room temperature results obtained after annealing the initially strained samples above their glass transition temperature. The average value of these is indicated by the dotted line. The solid lines in (a) and (b) are best fits to Eq.~\ref{energy_density}, with $k=2.3 \pm 0.5$ and $\gamma_\mathrm{sv} ^{(0)}-\gamma_\mathrm{sl} ^{(0)} = 47.5 \pm 0.1$~mJ m$^{-2}$, as well as $k=1.4 \pm 0.5$ and $\gamma_\mathrm{sv} ^{(0)}-\gamma_\mathrm{sl} ^{(0)} = 44.4 \pm 0.3$~mJ m$^{-2}$, respectively. Contact angle measurements are repeated several times at each strain, and the vertical error bars represent standard errors in these measurements. Uncertainties in the fitting parameters represent the 95\% confidence bounds.}
\label{fig3}
\end{figure}

In Figs.~\ref{fig3}(a) and~\ref{fig3}(b), we plot $\gamma_\mathrm{sv}-\gamma_\mathrm{sl}$, obtained via Young-Dupr\'e's law, as a function of strain for PSf and PC respectively, with DIM as the test liquid (circles). In both these cases, the contact angle increases with strain, causing $\gamma_\mathrm{sv}-\gamma_\mathrm{sl}$ to decrease. This result demonstrates, for the first time, the existence of strain dependent surface energies for interfaces involving a polymeric glass.

Using Eq.~\ref{Shuttleworth}, the surface stress difference at zero strain $\Upsilon_\mathrm{sv} ^{(0)}-\Upsilon_\mathrm{sl} ^{(0)}$, where the $(0)$ superscript will henceforth refer to the unstrained ($\epsilon=0$) case, can be determined by fitting a line to each of the data sets (circles) in Figs.~\ref{fig3}(a) and~\ref{fig3}(b). In doing so, the \emph{surface stress difference} at zero strain is found to be respectively 11~mJ m$^{-2}$  and 17~mJ m$^{-2}$ smaller than the \emph{surface energy difference} at zero strain $\gamma_\mathrm{sv} ^{(0)}-\gamma_\mathrm{sl} ^{(0)}$, for PSf and PC respectively. Similarly, the dependence of $\gamma_\mathrm{sv}-\gamma_\mathrm{sl}$ upon strain for PSf and PC with glycerol as the test liquid is shown (circles) in Fig.~\ref{fig3}(c) and~\ref{fig3}(d). A clear strain dependence is again observed. Surprisingly, $\gamma_\mathrm{sv}-\gamma_\mathrm{sl}$ increases with strain with this polar liquid, contrary to the results with DIM. Due to curvature in these data, they are not well described by a linear relationship, but fitting a line to the first few data points allows us to determine approximate values for $\Upsilon_\mathrm{sv} ^{(0)}-\Upsilon_\mathrm{sl}^{(0)}$.  The magnitude of the effect is much larger with glycerol as the test liquid, as $\Upsilon_\mathrm{sv} ^{(0)}-\Upsilon_\mathrm{sl}^{(0)}$  is found to be roughly 60~mJ m$^{-2}$ and 140~mJ m$^{-2}$ larger than $\gamma_\mathrm{sv} ^{(0)}-\gamma_\mathrm{sl}^{(0)}$, for PSf and PC respectively. 

After performing the measurements above using the strained glassy films (supported on silicon wafers), we anneal the supported films in the melt state ($\sim 30\,^\circ$C above their glass transition temperature) for 15 min. This procedure relaxes the pre-applied strain in the films. Next, the films are quenched back down to room temperature, and this thermal change induces a contraction. The expansion coefficient of silicon is much smaller than that of the polymer films.  Therefore, as the films re-enter the glassy state, their strong adhesion to the silicon wafers restricts them from contracting in the $x-y$ plane any further than the thermal contraction of the silicon. Therefore, although the pre-applied strain is erased by the annealing, a small biaxial strain $\epsilon_\mathrm{t} \approx 1\,\%$ is introduced due to thermal contraction~\cite{Fortais2017}. Then, we perform contact angle measurements once more, and the corresponding surface energy differences are plotted as squares in Fig.~\ref{fig3}. As can be seen in these plots, the value of $\gamma_\mathrm{sv}-\gamma_\mathrm{sl}$ is now constant with respect to the pre-applied strain $\epsilon$. These data intersect the previous measurements (circles) at a small non-zero strain, as expected from the small, $\epsilon_\mathrm{t}\approx 1\,\%$ biaxial strain present in the film due to thermal contraction.

To further assure that our results are not an artefact due to a plastic deformation of the surface upon straining, we performed two additional tests. First, atomic force microscopy scans (not shown) reveal no noteworthy difference between strained and unstrained films. No signs of crazing or anisotropic topography are seen on the strained sample, and the typical surface roughness is unchanged. Secondly, one PC film was strained to $\epsilon =$ 6\,\% -- below plastic yield -- and subsequently peeled off the holder blocks (Fig.~\ref{fig1}). The peeled film could thus relax the elastic deformation it was initially subjected to and was subsequently transferred onto a silicon wafer. The measured contact angle on that sample was found to lie within error of the value measured on an unstrained PC film.

A remaining question concerns the origin of the sign change between the slopes of the data sets for the two test liquids (Figs.~\ref{fig3}(a,b) and~\ref{fig3}(c,d) (circles)). The striking difference between the two liquids is that DIM (Figs.~\ref{fig3}(a,b)) is non-polar whereas glycerol (Figs.~\ref{fig3}(c,d)) is highly polar. To provide further evidence for the important role of polarity, we perform contact angle measurements with water on PSf. For the droplet sizes relevant to our experiments, the evaporation rate of water is too high to perform robust measurements of $\theta_\mathrm{Y}$. Instead, we perform advancing and receding contact angle measurements.  The advancing and receding contact angles of water on PSf  decreased by 7 $\pm$ 4$\,^\circ$ and 11 $\pm$ 3$\,^\circ$ over a 7\,\% strain, implying an increase in $\gamma_\mathrm{sv}-\gamma_\mathrm{sl}$. Thus, both polar liquids, water and glycerol, show the same increasing trend of $\gamma_\mathrm{sv}-\gamma_\mathrm{sl}$ with strain. Moreover, since surface energies characterize the molecular interactions at the interface, we would anticipate a significant difference whether these interactions are permanent - permanent dipole (Keesom force), permanent - induced dipole (Debye force), or induced - induced dipole (London dispersion force) in origin~\cite{Israelachvili2011}. Therefore, we suspect that the polarity of the liquid is the source of the difference in slope sign.

\subsection{Minimal Model}

The surface energy difference $\gamma_\mathrm{sv}-\gamma_\mathrm{sl}$ can be re-written as: $\gamma_\mathrm{sv}-\gamma_\mathrm{sl} = -\gamma_\mathrm{lv} + W_\mathrm{lvs}$, where $W_\mathrm{lvs}(\epsilon)$ is the work of adhesion between the liquid and solid (with vapour in between) and depends on strain. A simple treatment of the work of adhesion between two non-polar materials requires that the van der Waals interaction energy between two atoms is integrated for all pairs across the interface (Hamaker's calculation)~\cite{Israelachvili2011}. If we consider such a calculation for two semi-infinite half spaces of liquid and solid, $W_\mathrm{lvs}$ is proportional to the mass density product $\rho_\mathrm{l}\rho_\mathrm{s}$. As a simplification, we suppose that physical properties (\textit{e.g.} polarizability) other than $\rho_\mathrm{s}$ do not vary with strain. In this approach, a positive strain $\epsilon$ can induce a reduction in the density $\rho_\mathrm{s}$, which causes a proportional reduction in $W_\mathrm{lvs}$, implying a reduction in $\gamma_\mathrm{sv}-\gamma_\mathrm{sl}$. Indeed, the mass density of our films upon straining is given by $\rho_\mathrm{s}=\rho_\mathrm{s}^{(0)}[1-(1-2\nu)\epsilon/k]$ in the limit of small strains, where $\nu$ is the Poisson ratio of the film, and where the parameter $k$ depends on details of the straining geometry but is expected to be unity when the strained solid is completely unclamped at its sides in the $y$- and $z$-directions (Fig.~\ref{fig1}) but smaller than unity if the film is fully clamped. The constant $k$ is left as a free parameter in this minimal approach. Therefore, for dispersive interactions, we have a simple prediction for the strain dependence of the surface energy difference:
\begin{equation}
\gamma_\mathrm{sv}-\gamma_\mathrm{sl} = \gamma_\mathrm{sv}^{(0)} -\gamma_\mathrm{sl}^{(0)} - \left(\gamma_\mathrm{sv}^{(0)} -\gamma_\mathrm{sl}^{(0)}+\gamma_\mathrm{lv}\right)\frac{(1-2\nu)}{k}\epsilon.
\label{energy_density}
\end{equation}
Given that $\nu = 0.37$ for both PSf and PC~\cite{Margolis1985}, we can fit Eq.~\ref{energy_density} to our DIM/PSf and DIM/PC data leaving both $\gamma_\mathrm{sv}^{(0)} -\gamma_\mathrm{sl}^{(0)}$ and $k$ free. The results are shown as solid lines in Fig.~\ref{fig3}(a) and Fig.~\ref{fig3}(b). These fits describe the data well, and from these we extract values of $\gamma_\mathrm{sv}^{(0)} -\gamma_\mathrm{sl}^{(0)}$ which are in agreement with those obtained from contact angle measurements on unstrained films (see Fig.~\ref{fig3}(a,b)), and determine $k$ to be $2.3 \pm 0.5$ and $1.4 \pm 0.5$ for DIM/PSf and DIM/PC respectively. Though $k$ is of order unity, as expected, the minimal model is missing some important ingredients. For instance, the polarizability of the molecules in the solid may change with strain, or the surface density may behave differently under strain compared to the bulk. The simple model we have proposed is applicable to the dispersive interactions between a non-polar liquid and a solid, but cannot be simply extended to interactions involving permanent dipoles. Indeed, a polar liquid like glycerol introduces an additional degree of complexity in the interfacial interactions~\cite{Ibach1997}.

\subsection{Measurements on Elastomers}

\begin{figure}[t!]
     \includegraphics[width=\columnwidth]{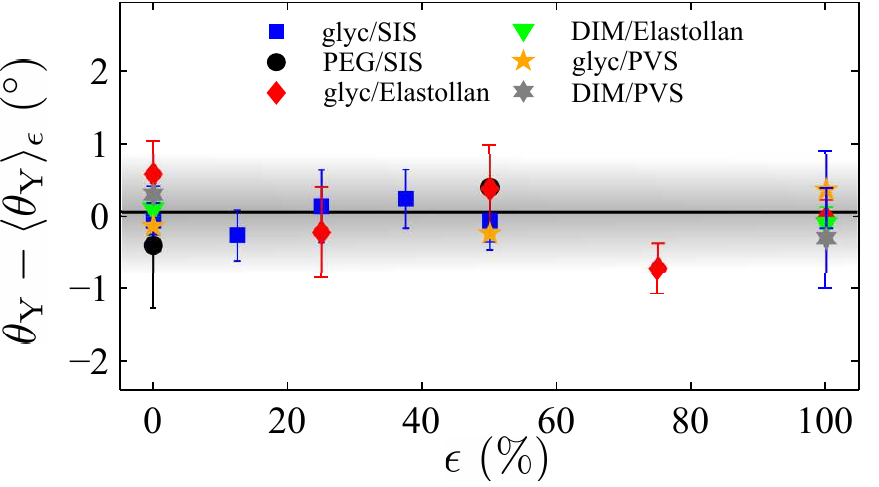}
\caption{Equilibrium contact angles relative to their average over all strains as a function of strain, for three elastomers using three different test liquids. Equilibrium contact angles are compared to the average value over all strains because this provides better statistics for the normalization compared to plotting with respect to the value at $\epsilon = 0$, in which case the normalization is determined only by one data point in each set and, as such, is more prone to error.   Vertical error bars represent standard errors in the measurement.}
\label{fig4}
\end{figure}

In the second part of this study, we perform contact angle measurements upon various elastomers using several test liquids. We employ two physically crosslinked elastomers: styrene-isoprene-styrene triblock copolymer (SIS), and Elastollan which is a thermoplastic polyurethane multiblock copolymer; as well as one chemically crosslinked elastomer: polyvinyl siloxane (PVS). We measure $\theta_\mathrm{Y}$ for these three elastomers using glycerol and DIM as the test liquids, with the exception of SIS for which we replace DIM by polyethylene glycol (PEG), since SIS is swollen by DIM. The results of the contact angle measurements for all liquid/elastomer combinations are shown in Fig.~\ref{fig4}, where we plot $\theta_\mathrm{Y} - \langle \theta_\mathrm{Y} \rangle _\epsilon$, \textit{i.e.} the deviation of the equilibrium contact angle from its mean value taken over all measured strains, as a function of strain. As seen in this plot, all contact angles remain constant within $\pm 1\,^\circ$ up to 100\,\% strain. These trends together with Young-Dupr\'e's law imply that $\frac{\mathrm{d} \gamma_\mathrm{sv}}{\mathrm{d} \epsilon} = \frac{\mathrm{d} \gamma_\mathrm{sl_1}}{\mathrm{d} \epsilon} = \frac{\mathrm{d} \gamma_\mathrm{sl_2}}{\mathrm{d} \epsilon}$ for all strains, where 1,2 indicate the two different test liquids. However, there is no physically sound reason to expect the solid-vapour surface energy to change by a non-zero amount under strain in exactly the same way as the solid-liquid surface energy, for an arbitrary choice of test liquid. In fact, one might expect the polarity of the liquid to play an important role. Thus, a reasonable expectation is that $\frac{\mathrm{d} \gamma_\mathrm{sv}}{\mathrm{d} \epsilon} = \frac{\mathrm{d} \gamma_\mathrm{sl}}{\mathrm{d} \epsilon} = 0$ for the interfaces involving the elastomers, which would imply through Eq.~\ref{Shuttleworth} that $\Upsilon_\mathrm{sv} = \Upsilon_\mathrm{sv}^{(0)}=\gamma_\mathrm{sv}=\gamma_\mathrm{sv}^{(0)}$ and $\Upsilon_\mathrm{sl} = \Upsilon_\mathrm{sl}^{(0)}=\gamma_\mathrm{sl}=\gamma_\mathrm{sl}^{(0)}$, and thus no Shuttleworth effect. Since we have tested several elastomers (physically and chemically crosslinked) and liquids (with varying polarity), we conjecture that this suggested property is applicable to solid-fluid interfaces involving elastomers in general.  If correct, this conjecture may be understood in the following simple way: elastomers are essentially incompressible ($\nu \approx 0.5$) liquids which are constrained by crosslinks on length scales much larger than those relevant to intermolecular interactions. Thus -- despite the strain -- the local molecular environment, density, and consequently stress and energy near the interface remain mostly unchanged. 

\section{Discussion}
In carrying out contact angle measurements, care must be taken to ensure that contact angle hysteresis does not cause artefacts. Here, the contact angle hysteresis is small (e.g. $<5^\circ$ for glycerol on PSf) on the glassy substrates since the spincoated films are uniform and clean. We find the measured contact angle to be highly reproducible from one droplet to the next. Given the methods of droplet deposition employed in this study, the measured static contact angle is expected to be closer to the advancing contact angle, but is a reliable approximation of the true Young’s angle due to the small hysteresis present. Given all the consistency checks that have been performed, including advancing/receding contact angle measurements with water which exhibit a consistent trend with strain, it is clear that hysteresis cannot be the cause of our main observations.

The fact that droplets in these experiments are completely circular when viewed from above emphasizes an important point. We apply uniaxial strains and hence the surface stresses, which are tensor quantities, are different in the directions normal and tangential to the strain. However, since our droplets are orders of magnitude larger than the elastocapillary length, the macroscopic contact angles measured are determined by Young-Dupr\'{e}'s law and thus functions of the surface energies, which are scalar quantities~\cite{Style2013a}. For this reason, the macroscopic contact angle is constant around the circumference of the droplet and the droplet's shape is a spherical cap.

Our study is motivated by the on-going debate on whether or not surface stresses in elastomers are identical to surface energies~\cite{Marchand2012,Nadermann2013,Style2013a,Weijs2014,Park2014,Mondal2015,Andreotti2016a,Andreotti2016,Xu2016,Xu2017,Xu2017b,Liang2018}. One set of experiments measured the surface stresses of an interface involving PVS by dipping a rod of this elastomer into an ethanol bath and measuring the deformation of the rod both above and beneath the liquid-air interface~\cite{Marchand2012,Andreotti2016}. In the aforementioned experiment, $\frac{\mathrm{d} \gamma_\mathrm{sl}}{\mathrm{d} \epsilon} - \frac{\mathrm{d} \gamma_\mathrm{sv}}{\mathrm{d} \epsilon} = 43 \pm 10$ mN m$^{-1}$, in contradiction to our data which suggests that $\frac{\mathrm{d} \gamma_\mathrm{sl}}{\mathrm{d} \epsilon} - \frac{\mathrm{d} \gamma_\mathrm{sv}}{\mathrm{d} \epsilon} = 0$ for all solid-fluid interfaces involving elastomers (including PVS). However, the result of Refs.~\cite{Marchand2012,Andreotti2016} relies on a model of the system which has recently been questioned~\cite{Hui2016}. In addition, the measurements of the local strains are highly sensitive to any minute swelling the submerging liquid may induce, since swelling affects $\rho_\mathrm{s}$. 

Other experiments have utilized the shape of the wetting ridge created at the contact line of a liquid droplet on a soft PDMS substrate to determine surface stresses of interfaces involving an elastomer~\cite{Style2013a,Park2014}. In a recent study, PDMS substrates were strained up to 25\,\% and, thereafter, the wetting ridge was imaged to deduce the strain dependent surface stresses~\cite{Xu2017,Xu2017b}. It was found that $\Upsilon_\mathrm{sv}$ and  $\Upsilon_\mathrm{sl }$ -- approximated to be equal due to a specific choice of liquid -- more than doubled over the range of strains tested, despite the equilibrium contact angle remaining unchanged. Although this result is in principle consistent with our elastomer data in Fig.~\ref{fig4}, it is inconsistent with our suggestion that $\frac{\mathrm{d} \gamma_\mathrm{sv}}{\mathrm{d} \epsilon} = \frac{\mathrm{d} \gamma_\mathrm{sl}}{\mathrm{d} \epsilon} = 0$ for elastomers. A noteworthy point is that considerations of how the elastic stress due to the strain affects the shape of the wetting ridge are not included in that study. In addition, although PDMS is a ubiquitous material which is important to study, it is also a challenging material as it is known to contain uncrosslinked chains which can migrate to the surface and act as a lubricant~\cite{Hourlier2017}, among other possible unexpected effects~\cite{Rivetti2017}. We acknowledge that the same effects may be present in PVS as well, and stress that PVS was only studied here to facilitate a direct comparison with Ref.~\cite{Marchand2012}.

Let us stress that all these studies, whether providing evidence for or against the equality between surface stress and surface energy for solid-fluid interfaces involving elastomers, are highly dependent upon a model to extract $\Upsilon$ from the raw data. 
In our work, we simply rely on Young-Dupr\'e's law to attain the results we present and, in turn, directly probe the surface stress difference $\Upsilon_\mathrm{sv}-\Upsilon_\mathrm{sl}$.

As discussed in the context of our minimal model, the fact that the value of the factor $k$ (see Eq.~\ref{energy_density}) falls outside of the expected range may indicate that the relevant physics is not described solely by the density changes upon strain. In fact, at large extensions, the orientation of chains, and thus the polarizability, will also be modified. Previous work has shown that orientation of chains in semicrystalline materials can induce contact angle changes for large strains ($\sim$ 100\%)~\cite{Good1971}. This type of effect could even play a role in some elastomers which are rich in double bonds. In such materials, if there is a sufficiently strong strain-induced birefringence, it is possible that a surface energy change will exist upon strain, despite the density remaining constant. We also expect that polarizability effects may be responsible for the different behaviours we observe for polar vs. non-polar liquids.

In this study, we have investigated the strain dependence of the solid-vapour and solid-liquid surface energies of interfaces involving amorphous materials, using contact angle measurements. The glassy materials tested show a significant change in $\gamma_\mathrm{sv}-\gamma_\mathrm{sl}$ with strain, which serves as a first demonstration of the Shuttleworth effect for glassy materials.  In addition, we show that changing the polarity of the test liquid  switches the sign of the strain dependence of $\gamma_\mathrm{sv}-\gamma_\mathrm{sl}$. In contrast, we show that $\gamma_\mathrm{sv}-\gamma_\mathrm{sl}$ remains constant for strains as large as 100\,\%  for several different elastomers, using various test liquids with different polarities. Our data are consistent with the notion that incompressible elastomers do not exhibit a Shuttleworth effect. 

\section{Methods}

\subsection{Polymer Details and Annealing Protocols}
Polysulfone (PSf) with number-averaged molecular weight $M_{\textrm{n}} =$ 22\,kg/mol (Sigma-Aldrich) is dissolved in cyclohexanone (Sigma-Aldrich, puriss p.a.$>$99.5\,\%). Polysulfone films are made with a thickness of $h \approx400$ nm. These films are annealed at 220$\,^\circ$C for 12 h. The re-annealing after contact angle measurement is done at 220$\,^\circ$C for 15 min. Poly(Bisphenol-A Carbonate) (PC) with $M_{\textrm{n}} =$ 22\,kg/mol (Polymer Source Inc.) and polydispersity index of 1.9 is dissolved in chloroform (Fisher Scientific, Optima grade). Polycarbonate films are made with a thickness of $h \approx1200$ nm. These films are annealed at 170$\,^\circ$C for 12 h. The re-annealing after contact angle measurement is done at 175$\,^\circ$C for 15 min. Styrene-isoprene-styrene (SIS) triblock copolymer (Sigma-Aldrich) with a 14\,\% styrene content is dissolved in toluene (Fisher Scientific, Optima grade). These films are made with a thickness of $h\approx1300$ nm and annealed at 110$\,^\circ$C for 10 min. Elastollan TPU 1185A (BASF) is dissolved in cyclohexanone (Sigma-Aldrich, puriss p.a.$>$99.5\,\%). These films are made with a thickness of $h\approx250$ nm and annealed at 100$\,^\circ$C for 90 min. Polyvinyl siloxane elastomer (PVS) is made by mixing base and catalyst (RTV EC00 Translucid) at a 1:1 ratio. These films are made with thicknesses on the order of several hundred microns.

\subsection{Sample Fabrication and Straining Protocol}
With the exception of the PVS samples, all films are prepared by spincoating out of solution. The samples are cast onto freshly cleaved mica substrates (Ted Pella Inc.) to produce uniform films. Subsequently, all samples (except PVS) are annealed to relax the polymer chains and remove any residual solvent that may be present within the sample. The glassy films are scored into $\sim$1 cm squares using a scalpel blade. The elastomeric samples are also divided into squares of $\sim$1 cm but rather using a cotton-tip applicator which is wetted with acetone. The films are then floated onto the surface of an ultrapure water bath (18.2 M$\Omega \cdot$cm, Pall, Cascada, LS) and subsequently picked up using a home built sample holder (Fig.~\ref{fig1}).  The PVS samples are made by depositing a drop of the catalyst-base mixture onto a freshly cleaved mica substrate and spreading it into a film using a clean glass pipette then leaving one hour for the elastomer to cure. The film is subsequently peeled off the mica and placed onto the sample holder. The initial gap between the two blocks of the sample holder (\textit{i.e.} the length of the film being strained, see Fig.~\ref{fig1}) was fixed at $l_0 =$ 4.5 mm. The water-air surface energy ensures that the films are taut (albeit at a strain $<< 1\, \%$  for the glasses and $<5\,\%$ for the elastomers) while floating and during the transfer onto the sample holder. After drying of the residual water from the floating process, one of the blocks of the sample holder is held in place while the other is attached to a translation stage (Newport MFA-CC, SMC100CC). The blocks are then un-fixed and the film is stretched along the $x$-axis. For the glassy films, the block is moved at a speed of 10 $\mu$m/s equivalent to a strain rate of 2.2$\cdot 10^{-3}$ s$^{-1}$. Performing the straining at 20 $\mu$m/s  produces no observable difference in the final results; however, as the speed is increased above 100 $\mu$m/s, we observe an increase in the likelihood of crazing. For the elastomeric films, the block is moved at a speed of 100 $\mu$m/s to reduce the time required to achieve the large strains. We observe no difference in the results when these films are strained at a speed of 10 $\mu$m/s. 

Once strained, the sample holder is rotated upside down and carefully translated down until the film bridging the gap between the two blocks makes contact with a silicon wafer which is placed below. Strong adhesive forces between the film and the smooth silicon wafer ensure that the film remains fixed to the wafer and thus unable to relax its strain. At this point, we cut the edges of the film with a scalpel to remove it from the sample holder, completing the sample preparation. The atomic force microscopy scans to ensure there are no apparent topographical changes to the strained surfaces in comparison to unstrained are performed with a Bruker, Multimode 8.

\subsection{Contact Angle Measurements}
Contact angle measurements are performed under an optical microscope. Measurements are performed immediately following droplet deposition. For diiodomethane (Sigma-Aldrich, Reagent Plus, 99\,\%), the contact radii $r$ of the droplets are in the range of $300<r<500$ $\mu$m. For glycerol (Caledon Laboratories Ltd.) and polyethylene glycol (Sigma-Aldrich, $M_{\textrm{n}}= 0.6$ kg/mol), the contact radii are in the range of  $50<r<350$ $\mu$m. The sufficiently large size of the droplets and the rapidity of the measurement ensure that evaporation does not significantly affect our measurements. For the PVS experiments, we work only with droplets with $r>$ 200 $\mu$m to ensure that the droplets are much larger than the elastocapillary length of the system. Droplets are placed as close as possible to the center of the film where the strain is least affected by the boundary conditions of the experimental straining geometry. However, we observe no systematic difference in the contact angles depending on the location of the droplet on the film. The droplet profiles are fit to circular caps to extract their radius of curvature $R$, from which the contact angle is attained using the relation $\mathrm{sin}(\theta_\mathrm{Y}) = r/R$. For each sample, several different droplets are imaged and the average contact angle is determined. 

To rule out potential effects due to swelling, we only employ liquid/solid combinations which are known to be highly immiscible. Since we work with thin films, which undergo colour changes upon minute thickness or refractive index changes, it is easily verified that there is no significant swelling of our films upon exposure to the test liquids used since colour changes are not observed. 

It is known that interfacial properties may depend upon the method with which the sample was prepared. For instance, a film which has been spincoated may exhibit different properties on the interface which was in contact with the substrate compared with the free interface. In order to be consistent, all our contact angle measurements are performed on the interface of the film which was in contact with the mica during spincoating.

\subsection{Data availability} 
The data that support the findings of this study are available from the corresponding author upon request.

\section{Acknowledgments}
The financial support by Natural Science and Engineering Research Council of Canada is gratefully acknowledged. The authors thank the Global Station for Soft Matter, a project of Global Institution for Collaborative Research and Education at Hokkaido University, as well as the Joliot chair from ESPCI Paris. We thank Y. Amarouchene, B. Andreotti, A. Antkowiak,  L. Bureau, A. Chateauminois, M. Chaudhury, K. Daniels, C. Drummond, E. Dufresne, A. Hourlier-Fargette, H. Perrin, P. Rambach, F. Restagno, J. Snoeijer, R. Style, and Q. Xu for valuable suggestions and comments.

\section{Author Contributions}
All authors contributed to designing the research project, R.D.S. performed all experiments and analyzed the data, M.T. lead the development of the theoretical model with input from all authors, R.D.S. wrote the first draft of the manuscript, and all authors edited the manuscript to generate a final version and contributed to the discussion throughout the entire process of the research.

\bibliography{Schulman2017bib}

\end{document}